\documentclass{article}
\usepackage{graphicx} 
\usepackage{authblk}

\usepackage[margin=1in]{geometry}
\DeclareUnicodeCharacter{2212}{\textminus}

\title{$\mathbf{3\times10^{-16}}$ \textbf{Stable Laser Deployed on a 4.3 km Mountain}}
\date{}

\author[1,2]{Tanner Grogan}
\author[1,2]{Eric Swiler}
\author[1,2]{Tristan Rojo}
\author[1]{Tobias Bothwell}
\author[1]{Robert J. Fasano}
\author[1]{Adam Halaoui}
\author[1]{Roger C. Brown}
\author[1,2,3,*]{Andrew D. Ludlow}

\affil[1]{National Institute of Standards and Technology, 325 Broadway, Boulder, Colorado 80305, USA}
\affil[2]{Department of Physics, University of Colorado, Boulder, Colorado 80309, USA}
\affil[3]{Electrical, Computer \& Energy Engineering, University of Colorado, Boulder, Colorado, USA}
\affil[*]{andrew.ludlow@nist.gov}

\begin{document}

\maketitle

\begin{abstract}
We present a transportable stable laser system based on a 10 cm reference cavity and an all-fiber phase-stabilized distribution network for the $\mathrm{^{171}Yb}$ clock wavelength. We highlight an extended cuboid cavity geometry with 4 times smaller spacer volume than a traditional cubic cavity of the same length \cite{Gill11}. We demonstrate repeatable acceleration sensitivities below $2\times10^{-10}g^{-1}$ in all axes and fractional frequency instability as low as $3\times10^{-16}$. We also demonstrate accurate distribution of the stable light over ~10 m fibers below the $1\times10^{-18}$ level in an all-fiber optic system. We report the successful deployment of the laser system to the summit of Mount Blue Sky at 4300 m, and for the first time, we report 16-digit laser instability measured outside the lab.
\end{abstract}

The past decade has seen rapid development in quantum and optical sensors that utilize state-of-the-art coherent light. In particular, optical atomic clocks have demonstrated uncertainty at the 18th digit and beyond, exceeding the current $\mathrm{^{133}Cs}$ standard \cite{Mcgr18,Aepp24,Jia25,Mars25}. One of the major obstacles to redefinition of the SI second is the need for clock comparisons across continents \cite{Dima24}. Metrologists around the world are building transportable optical clocks that can be shipped overseas towards SI redefinition and for chronometric leveling \cite{Taka20,Grot24,ICON24,Both25}. At the same time, the success of the GRACE Follow-On mission in positioning two satellites with nanometer precision using laser ranging interferometry \cite{Abic19} has inspired a wave of proposals for precision measurements in space. Both of these technologies rely on robust stable lasers to provide highly coherent probe light in harsh environments.

State-of-the-art reference cavities have demonstrated fractional frequency instabilities below $5\times10^{-17}$ at room temperature \cite{Ma25,Park26} and below $3\times10^{-17}$ at cryogenic temperatures \cite{Lee25,Yu23}. However, these cavities are loosely supported with precise geometry to minimize vibration sensitivity. Transportable cavities must be locked in place during transit while maintaining low vibration sensitivity \cite{Gill11,Chen14}. Additionally, transportable laser systems must take into account size, weight and power constraints without sacrificing too much on performance. Currently, the best transportable stable laser systems have demonstrated low 16-digit frequency stability in the laboratory \cite{Taka20,Herb22,Xiao22,Krau25} and have reported at least 15-digit stability in the field \cite{Grot24,ICON24,Cole24,Both25,Rosl24,Hilt25}.

We present a transportable stable laser system (TSLS) that generates 16-digit stable light at the $\mathrm{^{171}Yb}$ clock wavelength. The TSLS consists of a transportable cavity (CavT) with $3\times10^{-16}$ instability and a clock laser routing center (CLaRC) that disseminates cavity- or clock-referenced light with fiber noise cancellation (FNC) as accurately as $1\times10^{-18}$ at 5,000 s. We transported this system 105 km from our laboratory at NIST to the summit of Mount Blue Sky (4300 m), where we confirmed the laser stability using our transportable $\mathrm{^{171}Yb}$ lattice clock as a frequency discriminator. Additionally, the success of CavT validates a spacer design that emphasizes manufacturability without sacrificing repeatability.

\begin{figure}[ht]
\centering
\fbox{\includegraphics[width=\linewidth]{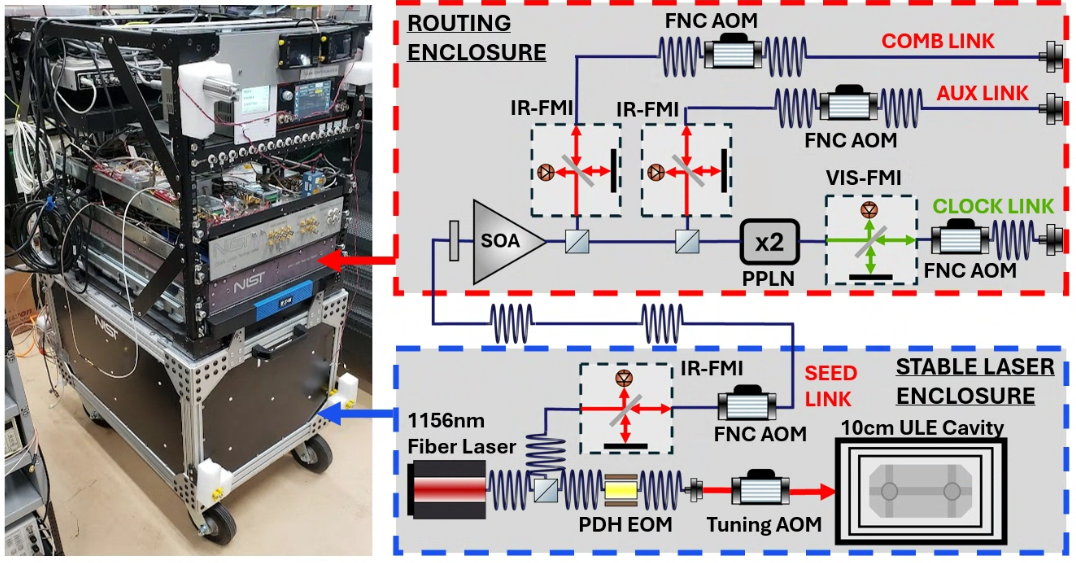}}
\caption{Schematic of the transportable stable laser system (right) and a photograph of the assembled mobile package (left). IR-FMI: infrared fiber-optic Michelson interferometer. VIS-FMI: visible fiber-optic Michelson interferometer. In the stable laser enclosure, an 1156 nm fiber laser is Pound-Drever-Hall (PDH) locked to CavT. Stable light is transferred to the routing enclosure through the "seed" FNC link. In the routing enclosure, the light is amplified to seed a periodically-poled $\mathrm{LiNbO_3}$ (PPLN) doubler which generates light at the $\mathrm{^{171}Yb}$ clock wavelength 578 nm. The "clock" FNC link delivers stable light to the atoms. Two 1156 nm FNC links ("comb" and "aux") are provided for flexible measurement configuration in the field.}
\label{fig:system}
\end{figure}

\begin{figure*}[ht]
\centering
\fbox{\includegraphics[width=\linewidth]{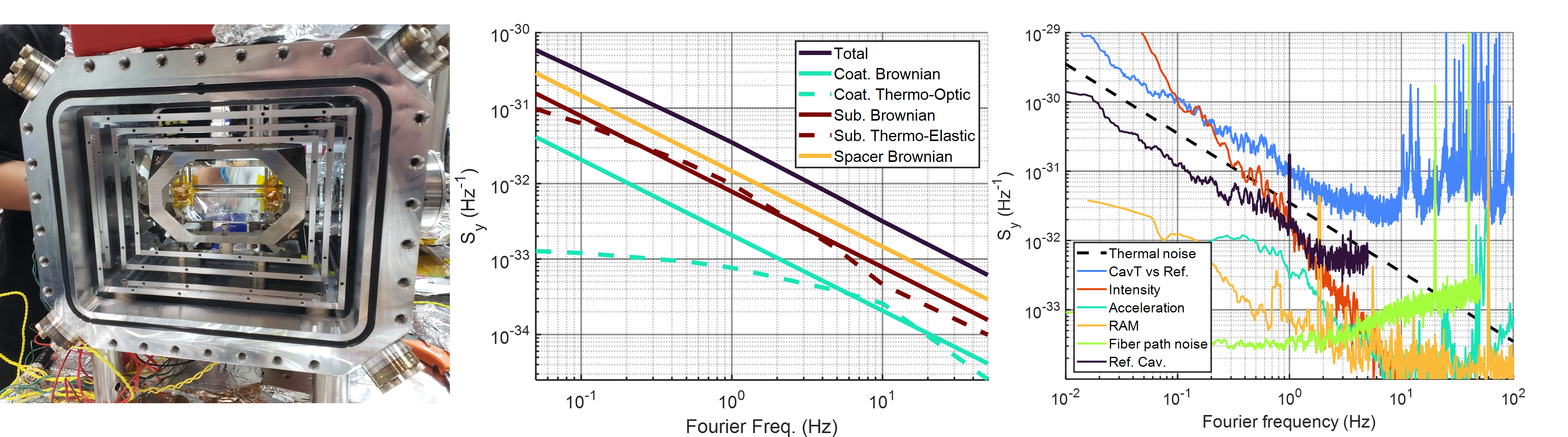}}
\caption{Left: CavT mounted in the cage structure with thermal shields. Middle: Calculated thermal noise budget for CavT. Right: Measured technical instability budget for CavT. Note the omission of spontaneous birefringent noise \cite{Ma26Laser} as there is not yet a quantitative theory for this noise process.}
\label{fig:cavity}
\end{figure*}

\begin{figure}[ht]
\centering
\fbox{\includegraphics[width=0.5\linewidth]{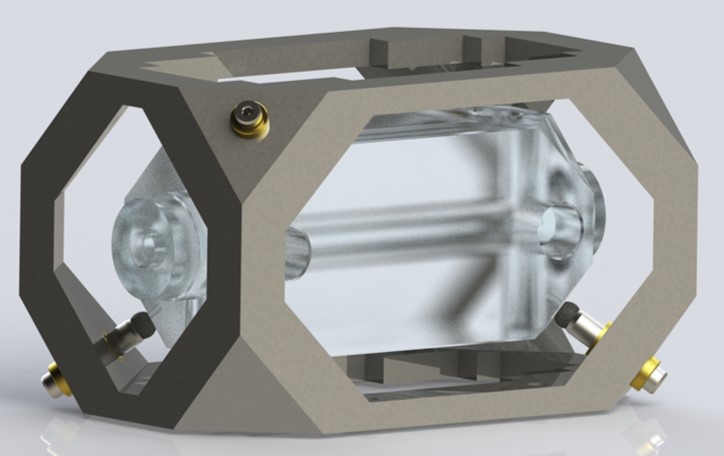}}
\caption{CavT geometry and mounting assembly. CavT is held in a stainless steel cage by four adjustment screws tipped with viton balls. Reference surfaces machined into the cage allow the cavity to be centered with gauge blocks during installation.}
\label{fig:cavgeometry}
\end{figure}

Our target for CavT was to achieve low 16-digit instability with a compact design that would be easily replicable in future iterations. One well-known obstacle to replicability is variation in the cavity's acceleration sensitivity each time the cavity is mounted to its support structure \cite{Leib13,Haef15}. To avoid this issue, we prioritize a simple mounting geometry and assembly process. Although the wire-mounted cavity developed in \cite{Herb22} has demonstrated superior and remarkably consistent acceleration sensitivities in the low $10^{-11}g^{-1}$'s in a small volume \cite{Herb22,Krau25}, mounting the cavity on wires (or threaded Invar rods in \cite{Xiao22}) involves some added complexity (wires, custom fasteners, arc-welds). 
On the other hand, the cubic cavity developed by Webster et al. \cite{Gill11} has only four holding screws, simplifying assembly at the cost of a larger spacer. Our extended cuboid design (Figure \ref{fig:cavgeometry}) inherits this simplicity while reducing the volume of the spacer by a factor of 4 for a given cavity length.

Following \cite{Gill11}, CavT's spacer geometry is optimized using finite-element analysis to make it insensitive to inertial forces and the applied holding force (see Supplement) \cite{Fasa21}. We choose viton as the material of the mounting ball to avoid plastic deformation. We repeatably achieve acceleration sensitivities on the order of $1\times10^{-10}g^{-1}$, as shown in Table \ref{AccelSens}. Accelerations in the stable laser enclosure are suppressed with an integrated active vibration isolation (AVI) platform. As reflected in our instability budget (Figure \ref{fig:cavity}, right), this is more than sufficient for low 16-digit performance. We attempted to further improve the acceleration sensitivity by adjusting the mounting screws \textit{in situ} with rotary feedthroughs for our vacuum chamber (see Supplement for details).

\begin{table}[htbp]
\centering
\caption{\bf CavT Acceleration Sensitivity ($10^{-10}g^{-1}$)}
\begin{tabular}{cccc}
\hline
 & $\mathrm{k_{axial}}$ & $\mathrm{k_{vertical}}$ & $\mathrm{k_{horizontal}}$ \\
\hline
1st time mounting & $0.98(.17)$ & $1.12(.15)$ & $0.41(.05)$ \\
2nd time mounting & $1.72(.11)$ & $0.50(.30)$ & --- \\
As deployed & $1.44(.10)$ & $0.60(.11)$ & $0.15(.02)$ \\
\hline
\end{tabular}
  \label{AccelSens}
\end{table}

To achieve low 16-digit instability in a small package, CavT is chosen to be 10 cm long. To minimize thermomechanical noise, we use $\mathrm{Al_{0.92}Ga_{0.08}As/GaAs}$ high reflective (HR) coatings on fused silica (FS) substrates with a 10.2 m radius of curvature. We measure a TEM$_{00}$ finesse of 253,000. Following \cite{Kess12}, we use finite element analysis (FEA) to estimate the Brownian motion of the ULE spacer. The thermal noise contributions are summarized in Figure \ref{fig:cavity} center (see Supplement for details). 

To stabilize the drift rate, CavT's spacer is machined from Corning 7973 ULE glass. We added 6.35 mm ULE backing rings to the mirrors to shift the zero-crossing of the coefficient of thermal expansion (ZCTE) to 30 $^\circ$C \cite{Lege10}. We measured the ZCTE near 20 $^\circ$C, and we discuss this discrepancy in the supplement. The heat dissipated by internal powered components raises the temperature in the stable laser enclosure to near 30 $^\circ$C. For practical operation, the cavity vacuum chamber is stabilized at 32 $^\circ$C using heaters. We rely on three layers of passive thermal shielding (Figure \ref{fig:cavity}) to reduce thermal fluctuations. The cavity drift rate is limited by thermal drifts typically around 100 mHz/s, canceled to first order by a phase-continuous frequency ramp applied to the RF reference of the seed FNC link (Figure \ref{fig:system}). By adjusting the compensation ramp in roughly 30 minute intervals, we are able to limit the uncompensated drift rate below 3 mHz/s.

\begin{figure*}[!ht]
\centering
\fbox{\includegraphics[width=\linewidth]{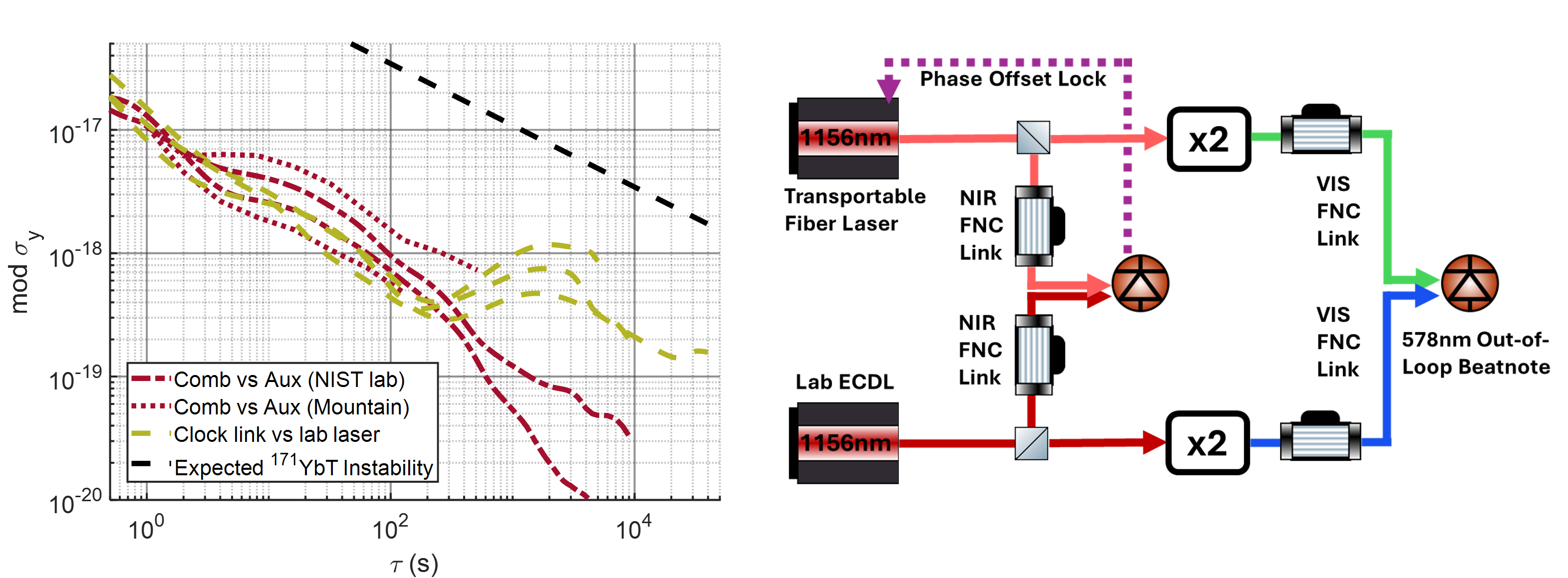}}
\caption{Left: Out-of-loop performance of the FNC links. The comb vs aux measurement is a direct stability measurement of the beatnote between these two links in a stable external free-space interferometer. Multiple identical measurements shown to highlight reproducibility.  Right: Experimental configuration for measuring phase fluctuations of the clock FNC link out-of-loop using the lab laser system. The CavT fiber laser is phase offset locked to the lab-based 1156 nm laser. The lab laser is doubled in free space, and that 578 nm light is sent through an FNC link to the out-of-loop beatnote with CLaRC's clock FNC link. We independently verified that the phase lock does not produce the rise in mod $\sigma_y$ for $\tau$>200 s.}
\label{fig:578meas}
\end{figure*}

The stability of CavT is measured against a stationary 29 cm ULE cavity at NIST ($\mathrm{mod}$ $\sigma_y = 1\times10^{-16}$ for $\tau>1$ s) \cite{Mcgr18}. At timescales of 1 Hz and below, the instability budget (Figure \ref{fig:cavity}) is dominated by intensity fluctuations. The intensity dependence arises primarily from the photo-birefringent response of the mirror coating, the photo-thermo-elastic response of the substrate, and the photo-thermo-refractive response of the mirror coating \cite{Krau25,Ma26}. We control residual amplitude modulation (RAM) passively by optimal choice of phase modulator and by temperature stabilization, as in \cite{Herb22}. We typically forego active RAM cancellation for simplicity. We achieve frequency stability of $3.0\times10^{-16}$ from ** Speculatively, we note that the spontaneous birefringence noise observed in \cite{Ma26Laser}, which contributed $5\times10^{-17}$ in a 48 cm cavity at 1 s, would contribute $2.4\times10^{-16}$ in our 10 cm cavity if all other variables were the same. This agrees very closely with our observed instability at 1 s.

Our design target for CLaRC was to disseminate stable light with $1\times10^{-18}$ accuracy or better with long-term reliability. We generate one channel of 578 nm light to interrogate an $^{171}\mathrm{Yb}$ optical lattice clock and two channels of 1156 nm light for frequency comparisons with other optical references. CLaRC is built entirely from polarization maintaining (PM) fiber optics, inspired by \cite{Ohma19,Cole24}. In contrast to free-space distribution systems \cite{Herb22,Both25}, fiber optic components do not require re-alignment after transit, and the choice of PM fiber helps preserve the output polarization of the FNC links despite temperature changes and mechanical disturbances. 

\begin{figure}[ht]
\centering
\fbox{\includegraphics[width=0.8\linewidth]{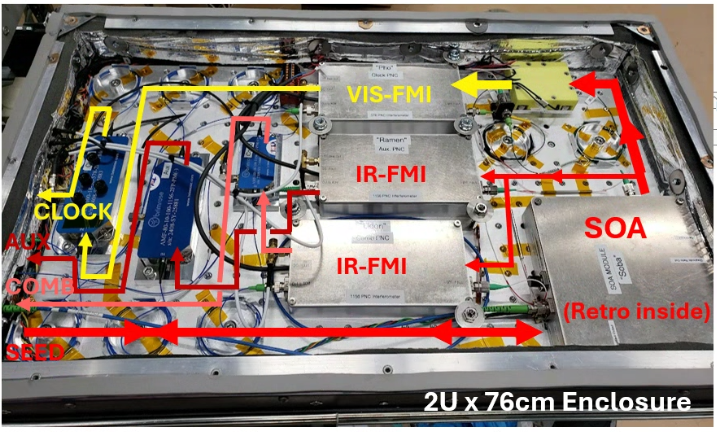}}
\caption{Physical and optical layout of CLaRC, located in the routing enclosure shown in Figure \ref{fig:system}.}
\label{fig:CLaRC}
\end{figure}

To further reduce environmental influence on the phase stability, we minimize all lengths of uncompensated fiber by fusion splicing. To house these uncompensated fiber optics, we engineer a 25 dB acoustically shielding enclosure with an internally temperature-stabilized breadboard with Peltier coolers, all contained within 2 rack units 76 cm deep (Figure \ref{fig:CLaRC}). For design flexibility as well as stability, we house each fiber optic Michelson interferometer in a gasket-sealed box. We also house the seed retroreflector, SOA, and necessary Faraday isolators in a gasket-sealed box to further shield the 7 m of uncompensated fiber within. 

In contrast to \cite{Ohma19}, we perform second harmonic generation in a fiber-coupled PPLN doubler. This has the advantage of eliminating the free-space doubler and steering optics on the clock physics package, which would require periodic re-alignment. Additionally, our architecture has the advantage that it compensates free space path fluctuations between the fiber launch on the physics package and the lattice retro-reflector of the clock. This mitigates the risk of apparent frequency shifts of the clock transition associated with cycle-synchronous vibrations on the physics package optics (for example, deflection of the optical path driven by magnetic fields switching on and off) \cite{Falk12}. Figure \ref{fig:578meas} shows the stability and accuracy of these links, comparable to \cite{Ohma19,Cole24}. We produce >4 mW at 1156 nm in the "comb" and "aux" links and 0.7 mW at 578 nm in the clock link, suitable for clock spectroscopy and longitudinal sideband cooling.

\begin{figure*}[ht]
\centering
\fbox{\includegraphics[width=\linewidth]{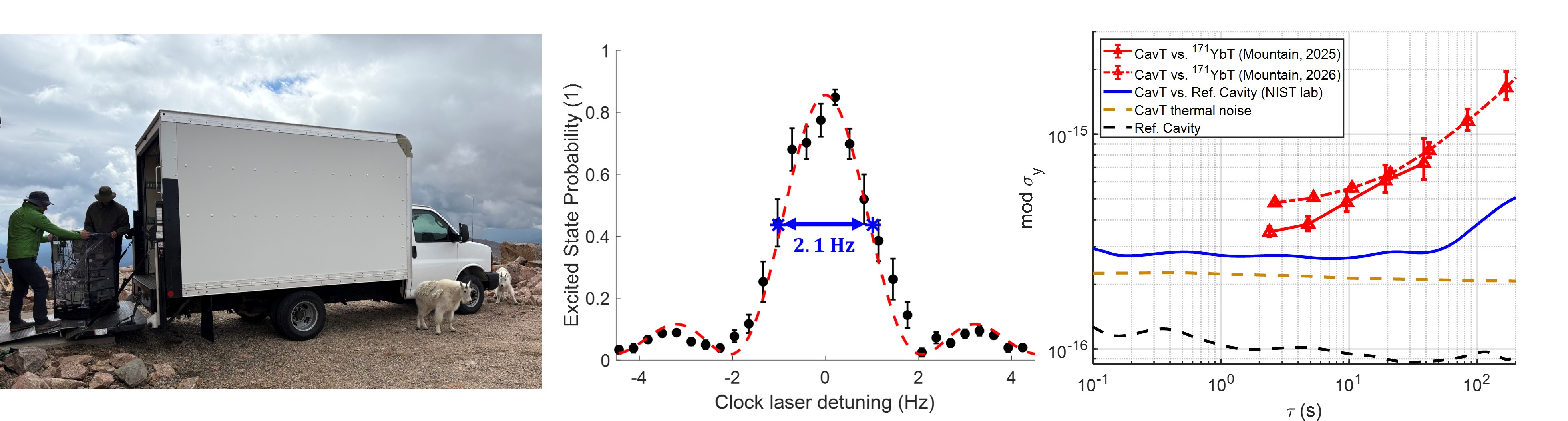}}
\caption{Left: The deployment team loading YbT onto the box truck to depart Mt. Blue-Sky. Center: Scan of the transportable laser through the YbT clock line with 418 ms Rabi probe time, acquired on Mount Blue Sky. Each point is the average of 18 shots with interleaved clock measurements for drift compensation. The dashed red line is an ideal Rabi lineshape with contrast fit to our data. Right: Stability measurements of CavT at NIST and on Mount Blue Sky with 240 ms Rabi probe time, after subtraction of a 5.9 mHz/s (2025) and 2.4 mHz/s (2026) residual drift. The residual drift reflects the difference between the cavity drift and the applied phase-continuous frequency ramp on the seed FNC RF reference. The fielded atomic side-of-line data reflects the instability of the cavity plus detection noise and the Dick effect at short times. The excess instability measured on the mountain at longer times is likely due to the less stable thermal environment.}
\label{fig:MtBS}
\end{figure*}

On August 8 2025, our transportable laser system was loaded on a box truck and traveled 105 km from NIST campus in Boulder, CO (1,655 m) to the Meyer-Womble observatory at the summit of Mount Blue Sky (4,326 m) (Figure \ref{fig:MtBS}). The observatory was not climate controlled. Inside the observatory, the deployment team erected a tent with a thermostat-controlled fan to protect the transportable clock system from excessive temperature changes. The temperature inside the tent still varied diurnally by $\pm5$ $^\circ$C. The stable laser package had no issues with the environment. Compensating CavT's thermal drift required human intervention about as frequently on the mountain as it did in lab. The system was transported to Mount Blue Sky again on June 11 2026, at which time we repeated our measurements from 2025. The changes made to the TSLS for the 2026 deployment are discussed in the supplement.

At the summit, we measured the stability of the TSLS using our transportable $\mathrm{^{171}Yb}$ optical lattice clock (YbT) as an independent frequency reference. We illuminate YbT with 240 ms Rabi pulses of 578 nm light from CLaRC's "clock" channel, with the laser's frequency detuned to the side of the line, where the fraction of the $\mathrm{^{171}Yb}$ atoms in the excited state is maximally sensitive to the TSLS frequency. With each pulse, we alternate between the rising and falling sides of the Rabi line (for a single Zeeman state). We estimate the TSLS frequency error relative to the atomic line by subtracting the excited state fraction measured on the falling side of the line from the excited state fraction measured on the rising side. To keep the TSLS from drifting off the side of the line during a long measurement, we apply a single frequency correction to the TSLS whenever the measured frequency error exceeds a quarter of the atomic linewidth. In the data presented in Figure \ref{fig:MtBS}, these automated corrections were applied roughly once per minute. To estimate the absolute frequency deviation of the TSLS, we add the measured frequency error relative to the atomic line to the accumulated automated frequency correction. We compute the modified Allan deviation of the measured laser frequency after subtracting a linear drift, shown in the right panel of Figure \ref{fig:MtBS}. Additionally, at the summit, we compare the "Aux" and "Comb" 1156 nm FNC links by combining the beams in free-space and counting the detected beatnote, as shown in Figure \ref{fig:578meas}. 

In conclusion, we have demonstrated a state-of-the-art TSLS compatible with $10^{-18}$-level optical lattice clock operation in the field. The stable reference cavity achieves $\mathrm{mod}\  \sigma_y=3\times10^{-16}$ at 1 s averaging times. The clock light routing network achieves $1\times10^{-18}$ routing accuracy in a compact and robust package without free-space optical components. Finally, we have demonstrated for the first time frequency stability at the low 16-digit level outside the laboratory in an extreme environment.

\textbf{Funding} This work was supported by NIST, DARPA, ONR, and NSF QLCI Award OMA-2016244.

\textbf{Acknowledgements} We acknowledge Theodora Triano, Grace Jenkins, William Swann, Ian Coddington, Marco Pomponio, Craig Nelson, and Archita Hati for technical contributions.

\textbf{Disclosures} The authors declare no conflicts of interest. No US
government endorsement of any company is suggested/implied in this
work.

\textbf{Data availability} Data underlying the results presented in this paper are available upon reasonable request.

See Supplement for supporting information.

\clearpage
\appendix
\part*{Supplementary Information}
\section{Cavity Thermal Noise}
Following \cite{Kess12}, the thermal noise of CavT can be broken down into Brownian thermomechanical contributions from the spacer, substrates, and mirror coatings, as well as a thermo-elastic contribution from the substrate and a thermo-refractive contribution from the coating. 
\begin{equation}
S_y(f) = 2S^{(coat)}_{y,Br}(f)+2S^{(sub)}_{y,Br}(f)+S^{(spac)}_{y,Br}(f)+2S^{(coat)}_{y,TR}(f)+2S^{(subs)}_{y,TE}(f)
\label{eq:S1}
\end{equation}
Following \cite{Kess12}, we estimate the 1/f coefficients of the coating and substrate Brownian contribution analytically (Figure 2, center panel). We also estimate the substrate thermo-elastic and coating thermo-refractive contributions analytically \cite{Herb22}.

\begin{table}[htbp]
\centering
\caption{\bf Material Properties for Thermal Noise Calculations at 32 $^{\circ}\mathrm{C}$ \cite{Herb22}}
\begin{tabular}{ccccc}
\hline
 & ULE & Fused silica & $\mathrm{Al_{0.92}Ga_{0.08}As/GaAs}$ & \\
\hline
$Y$ & 67.6 & 72 & 100 & Young's modulus (GPa) \\
$\phi$ & 16.7 & 1 & 0.24 & Loss angle ($10^{-6}$) \\
$\eta$ & 0.17 & 0.17 & 0.32 & Poisson ratio \\
$C$ & 1.7 & 1.64 & 1.70/1.75 & Volumetric heat capacity ($\mathrm{MJ\:m^{-3}\:K^{-1}}$) \\
$\alpha$ & 0.06(3) & 0.51 & 5.2/5.7 & Coefficient of thermal expansion ($10^{-6}$) \\
$\kappa$ & 1.38 & 1.38 & 70/55 & Thermal conductivity ($\mathrm{W\:m^{-1}\:K^{-1}})$ \\
$\beta$ & - & - & 179/366 & Effective thermorefractive index ($10^{-6}\:\mathrm{K^{-1}}$) \\
$n$ & - & - & 2.96/3.39 & Index of refraction \\
\hline
\end{tabular}
  \label{tab:matprops}
\end{table}

\begin{table}[htbp]
\centering
\caption{\bf Cavity Parameters for Thermal Noise Calculations}
\begin{tabular}{ccc}
\hline
Symbol & Value & Description\\
\hline
\hline
$\lambda$ & 1156 nm & Operating wavelength\\
$w$ & 513 $\mu\mathrm{m}$ & Spot size at mirror\\
\hline
$d_{subs}$ & 6.35 mm & Substrate thickness\\
$r_{subs}$ & 25.4 mm & Substrate outer diameter\\
$ROC$ & 10.2 m & Substrate radius of curvature\\
\hline
$d_{coat}$ & 6.6 $\mu \mathrm{m}$ & Coating thickness\\
\hline
$l$ & 10 cm & Spacer length \\
$F_h$ & 40 N & Holding force \\
\hline
\end{tabular}
  \label{tab:matprops2}
\end{table}

We follow the approach of \cite{Kess12} to estimate the Brownian noise of the spacer, accounting for its unique geometry. Using a finite element model, we apply a tensile pressure profile to each mirror surface that follows the TEM$_{00}$ beam profile:
\begin{equation}
p(r) = \pm \frac{2F}{\pi w^2}e^{-2r^2/w^2}
\label{eq:S2}
\end{equation}
where $F$ is a test force set to 1 N. We evaluate the change in the strain energy $U$ of the spacer in response to the modeled test force, and from $U$ we estimate the spacer Brownian noise contribution.
\begin{equation}
S^{(spac)}_{y,Br}(f) = \frac{4k_B T U \phi_{spac}}{\pi f F^2}
\label{eq:S3}
\end{equation}

The biggest contributor to our thermal noise estimate is the Brownian noise of the spacer, followed closely by the Brownian noise of the substrate (Figure 2, center panel). We evaluate the effect of the extended cubic spacer geometry by comparing the estimated Brownian spacer contribution of our spacer with that of a standard cylindrical geometry with similar dimensions (10 cm long, 5 cm diameter, same optical bore diameter, and identical mirror parameters). We find that the extended cubic spacer's contribution (1.20 nJ of elastic strain energy for a 1 N test force) is comparable to that of the cylindrical spacer (0.98 nJ). The slight enhancement in the extended cubic spacer appears to be correlated with the presence of the venting bores near the mirrors. The Brownian noise contribution of the extended cubic spacer would likely be reduced in a longer cavity, such as in \cite{Herb22,Xiao25,Krau25}.


\section{Force-Insensitive Extended Spacer}
\begin{figure}[htbp]
\centering
\fbox{\includegraphics[width=.7\linewidth]{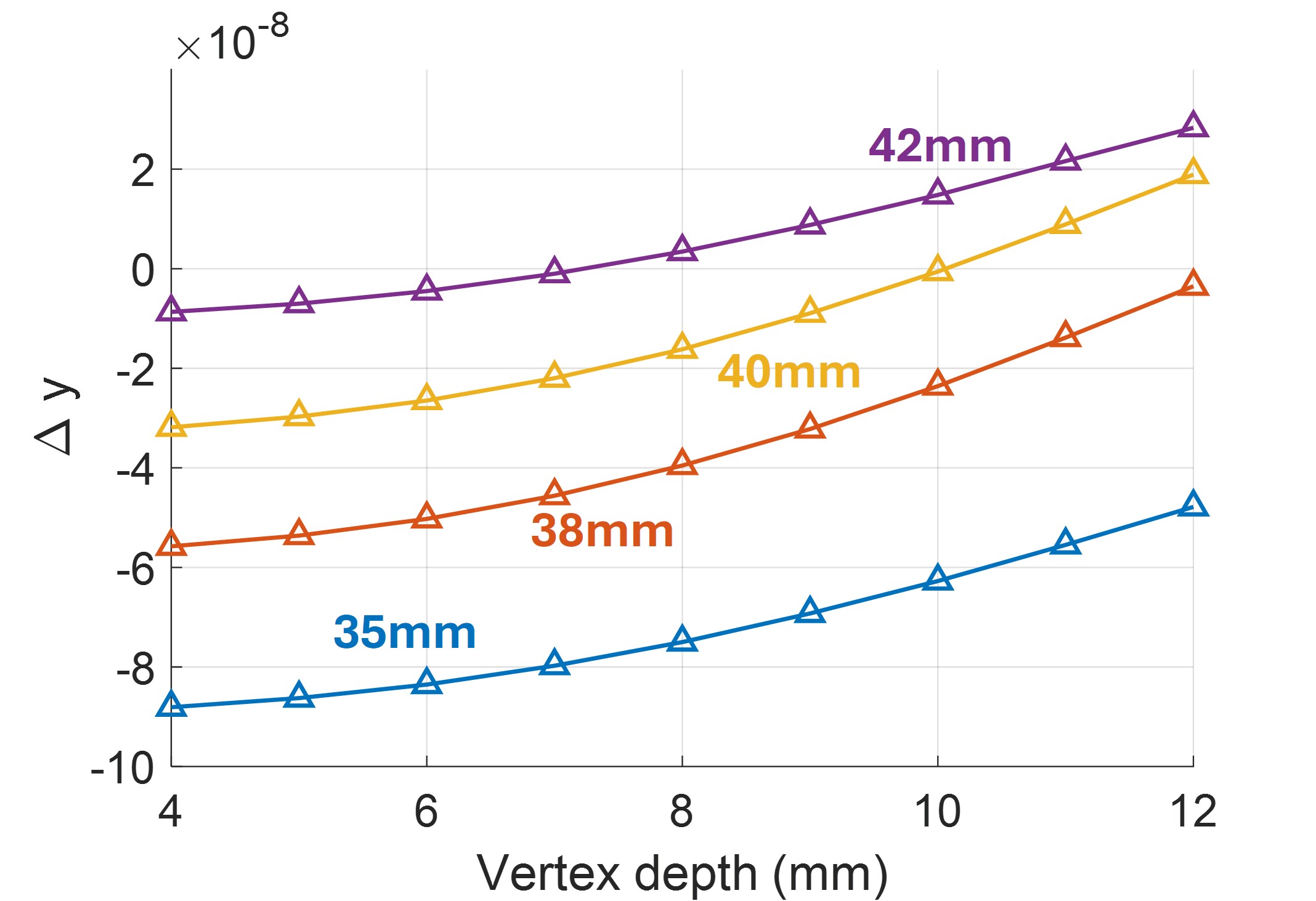}}
\caption{Force sensitivity study while varying the vertex depth. The different colors with associated dimensions in mm correspond to different values of the bore offset. The bore offset is defined as the distance from the center of a venting bore hole to the mid plane orthogonal to the optical axis. Equivalently, the bore offset is half the longitudinal distance between venting bore holes. In the final design presented in this work, the bore offset is 40 mm and the vertex depth is 10 mm.}
\label{fig:vdepth}
\end{figure}

To minimize the acceleration sensitivity of the cavity, the spacer geometry must be optimized using finite element analysis (FEA). The spacer geometry is inspired by NPL's cubic design \cite{Gill11}, and follows a similar engineering design process \cite{Fasa21}. In the design process of the NPL cubic cavity, the authors varied the vertex depth to cancel the sensitivity of the cavity length to the holding force (force sensitivity), and then they verified that the acceleration sensitivity was also canceled \cite{Gill11}. In our extended cuboid design, we find that extending the cavity along the optical axis shifts the force sensitivity negatively, and so we cannot cancel the force sensitivity using only the vertex depth. We additionally split the venting bores of the spacer into symmetric pairs of venting bores, and we vary the offset between venting bores as well as the vertex depth to minimize force sensitivity (Figure \ref{fig:vdepth}). As in the NPL case, we find that for the geometry optimized for force sensitivity, the acceleration sensitivity is also controlled at the $10^{-11}g^{-1}$ level.

We included four 1.5 mm blind holes on 4 corners of the spacer to align the spacer to its mounting cage during assembly using dowels. We found that using gauge blocks to center the spacer on the cage yielded more consistent results, so these 4 blind holes are unused. The other 4 corners are reserved for the mounting balls.

\begin{figure}[htbp]
\centering
\fbox{\includegraphics[width=.95\linewidth]{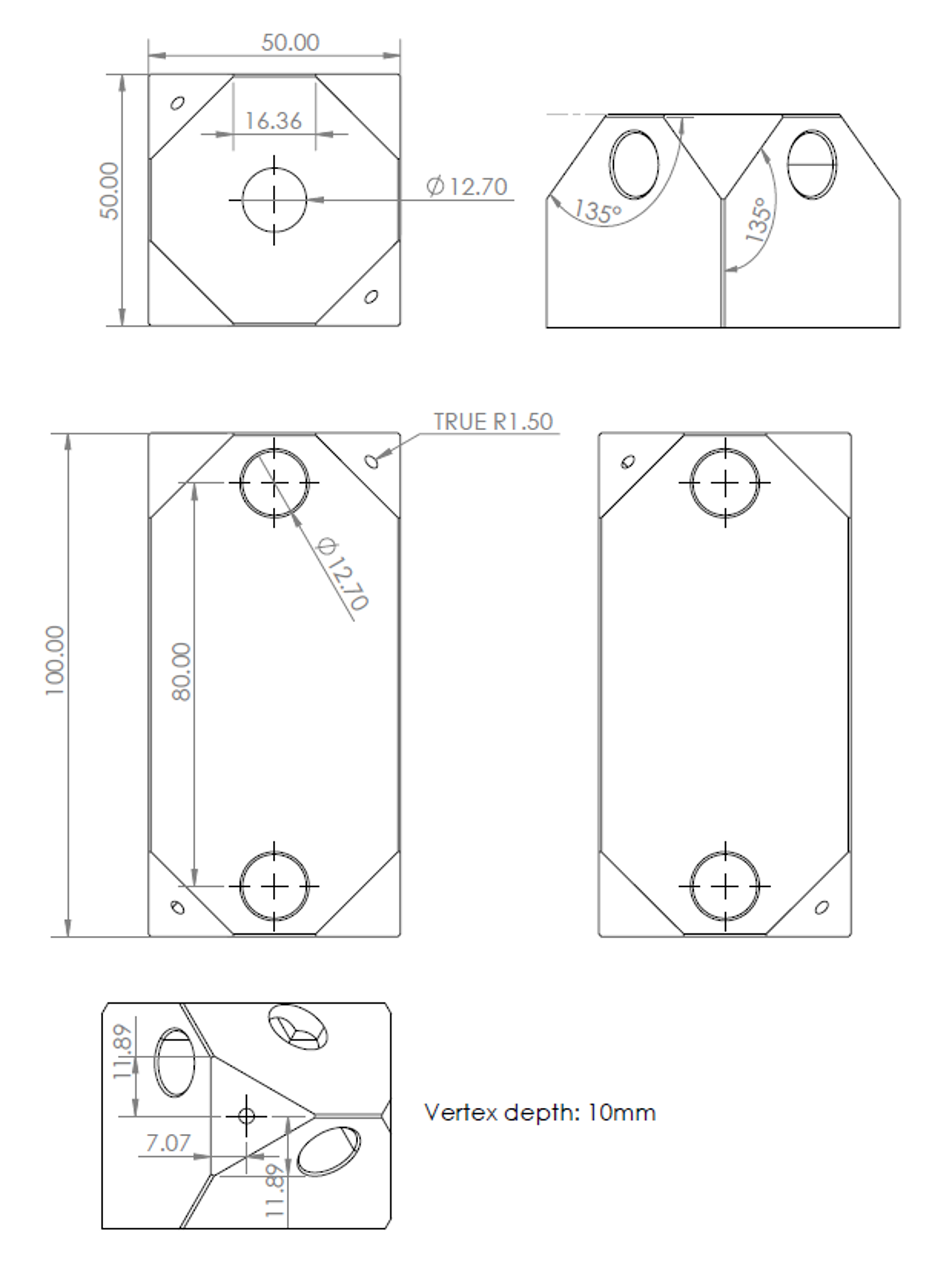}}
\caption{Mechanical drawing of CavT spacer. All dimensions are in millimeters. Vertex depth is defined perpendicular to the plane of the corner cut.}
\label{fig:drawing}
\end{figure}

A mechanical drawing of the spacer is shown in Figure \ref{fig:drawing}. We had all edges of the spacer (internal and external) chamfered to the best of the fabricators' ability to eliminate stress risers. Additionally, we had all internal surfaces of the spacer polished to reduce the generation of dust near the high finesse mirrors, and we had the end faces polished to allow optically contacting the mirror substrate. 

\section{Measurement of Acceleration Sensitivity}
\begin{figure}[htbp]
\centering
\fbox{\includegraphics[width=.95\linewidth]{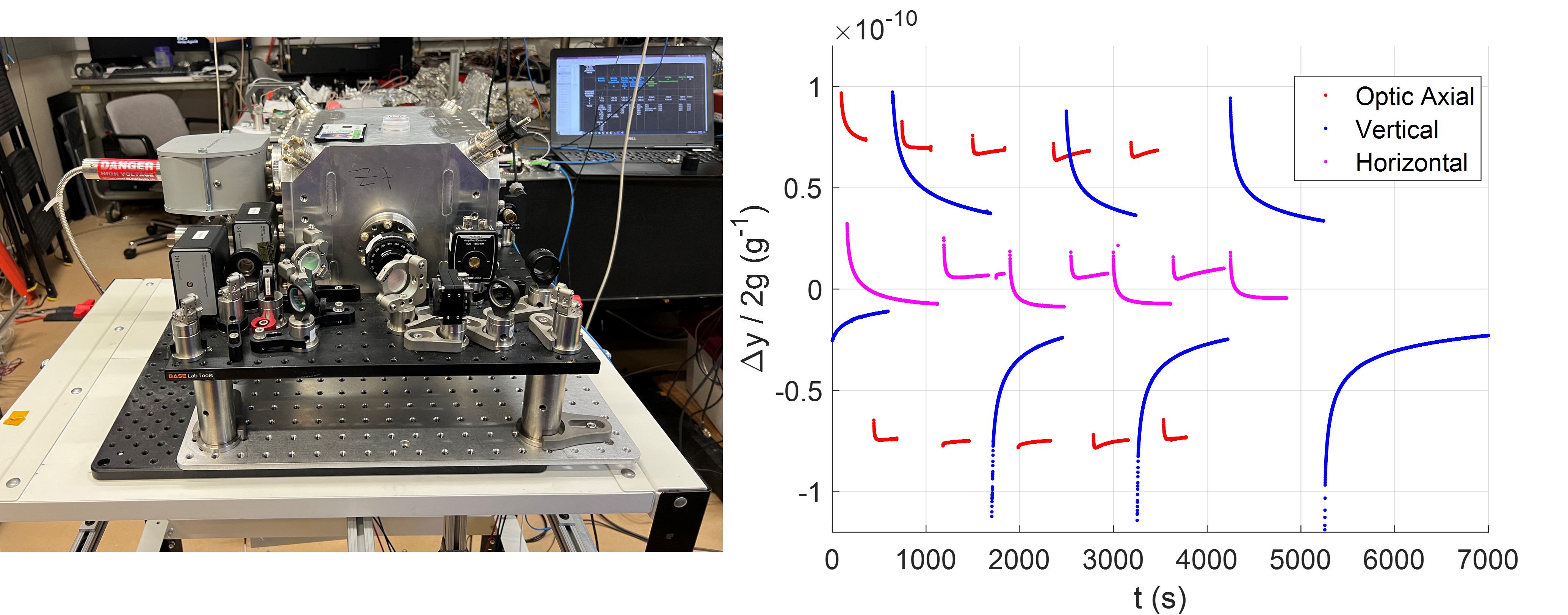}}
\caption{Left: Photograph of prototype optical setup on flip table for measuring CavT's acceleration sensitivity. Right: representative measurement of CavT acceleration sensitivity, showing CavT frequency measured against a stable optical reference while turning the flip table.}
\label{fig:fliptest}
\end{figure}

To measure the acceleration sensitivity, we measure the fractional frequency change when the cavity is flipped 180 degrees and divide by 2g. We perform these measurements using a prototype setup with a tabletop that rotates (Figure \ref{fig:fliptest}). The whole time CavT is held at vacuum below $1\times10^{-3}$ mPa by an ion pump. After each flip, we wait out any relaxation processes, such as photothermal relaxation after losing the CavT PDH lock durig a flip, before recording the frequency change.

\section{\textit{In Situ} Adjustment of the Acceleration Sensitivity}
\begin{figure}[htbp]
\centering
\fbox{\includegraphics[width=.95\linewidth]{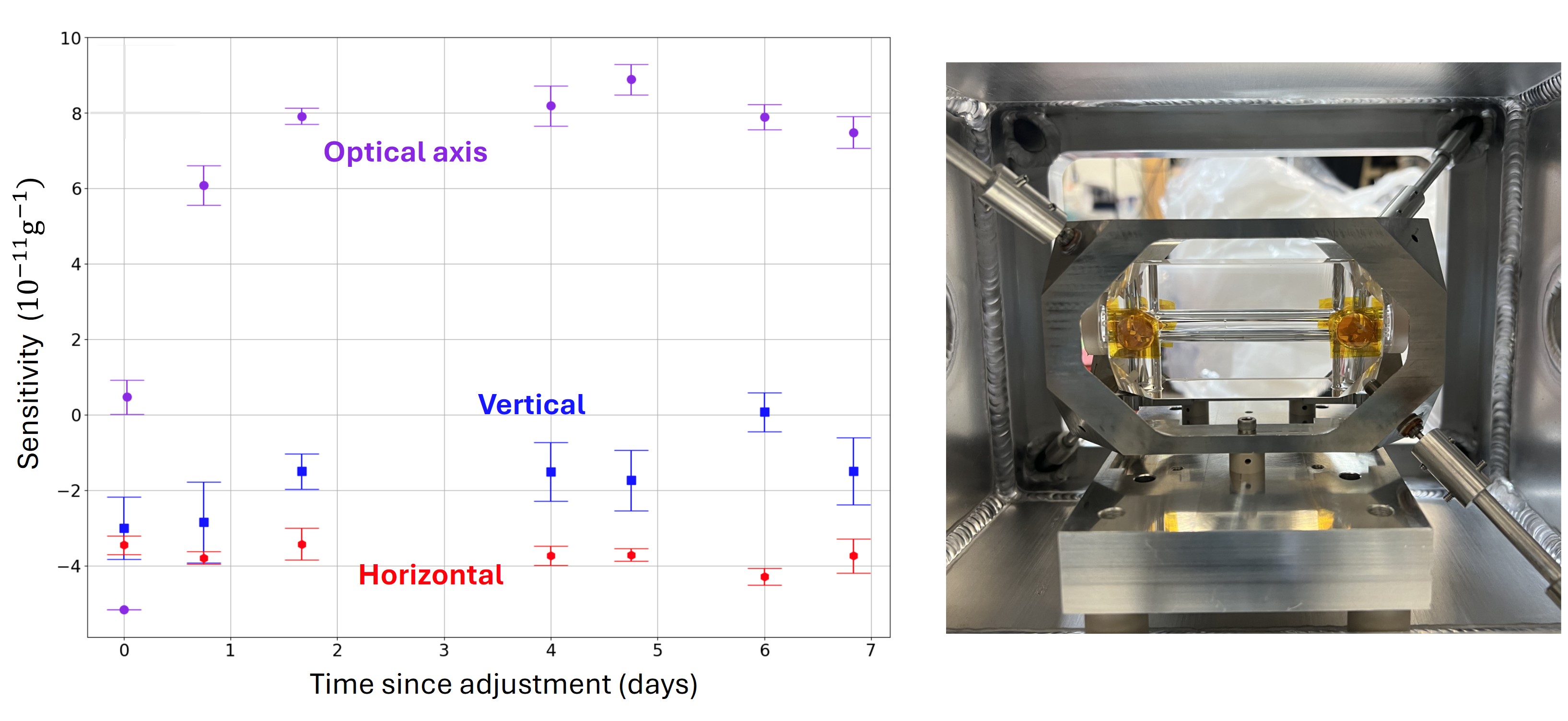}}
\caption{Left: Acceleration sensitivity of CavT over time after adjusting one of the mounting screws. Right: Photograph of CavT in vacuum chamber with in-vacuum coupling shafts to adjust mounting screws with vacuum-sealing rotary feedthroughs.}
\label{fig:feedthroughs}
\end{figure}

We attempted to further improve the acceleration sensitivity \textit{in situ} using rotary feedthroughs to turn the mounting screws without breaking vacuum. The vacuum chamber has four mini-conflat flanges for rotary feedthroughs, aligned with the cavity mounting screws, as shown in Figure \ref{fig:feedthroughs}.

After turning any of the screws, we observed a large time variation in the acceleration sensitivity, reaching a new steady state after several days (Figure \ref{fig:feedthroughs}). Because of the long delay between adjustment and measurement, \textit{in situ} compensation would take a long time. We speculate the day-scale settling behavior may be due to the mechanical relaxation of the viton ball that contacts the cavity. After attempting \textit{in situ} adjustment, we broke vacuum to transfer the mounting cage assembly into a separate vacuum chamber without feedthroughs for deployment. The effect of the transfer on acceleration sensitivity is summarized in Table \ref{CageTransfer}.
\begin{table}[htbp]
\centering
\caption{\bf CavT Acceleration Sensitivity Before and After Transferring Mounting Cage between Vacuum Chambers ($10^{-10}g^{-1}$)}
\begin{tabular}{cccc}
\hline
 & $\mathrm{k_{axial}}$ & $\mathrm{k_{vertical}}$ & $\mathrm{k_{horizontal}}$ \\
\hline
Before cage transfer & $0.92(.18)$ & $0.54(.24)$ & -- \\
After cage transfer & $1.44(.10)$ & $0.60(.11)$ & $0.15(.02)$ \\
\hline
\end{tabular}
  \label{CageTransfer}
\end{table}

\section{Thermal Expansion of CavT}
\begin{figure}[htbp]
\centering
\fbox{\includegraphics[width=0.99\linewidth]{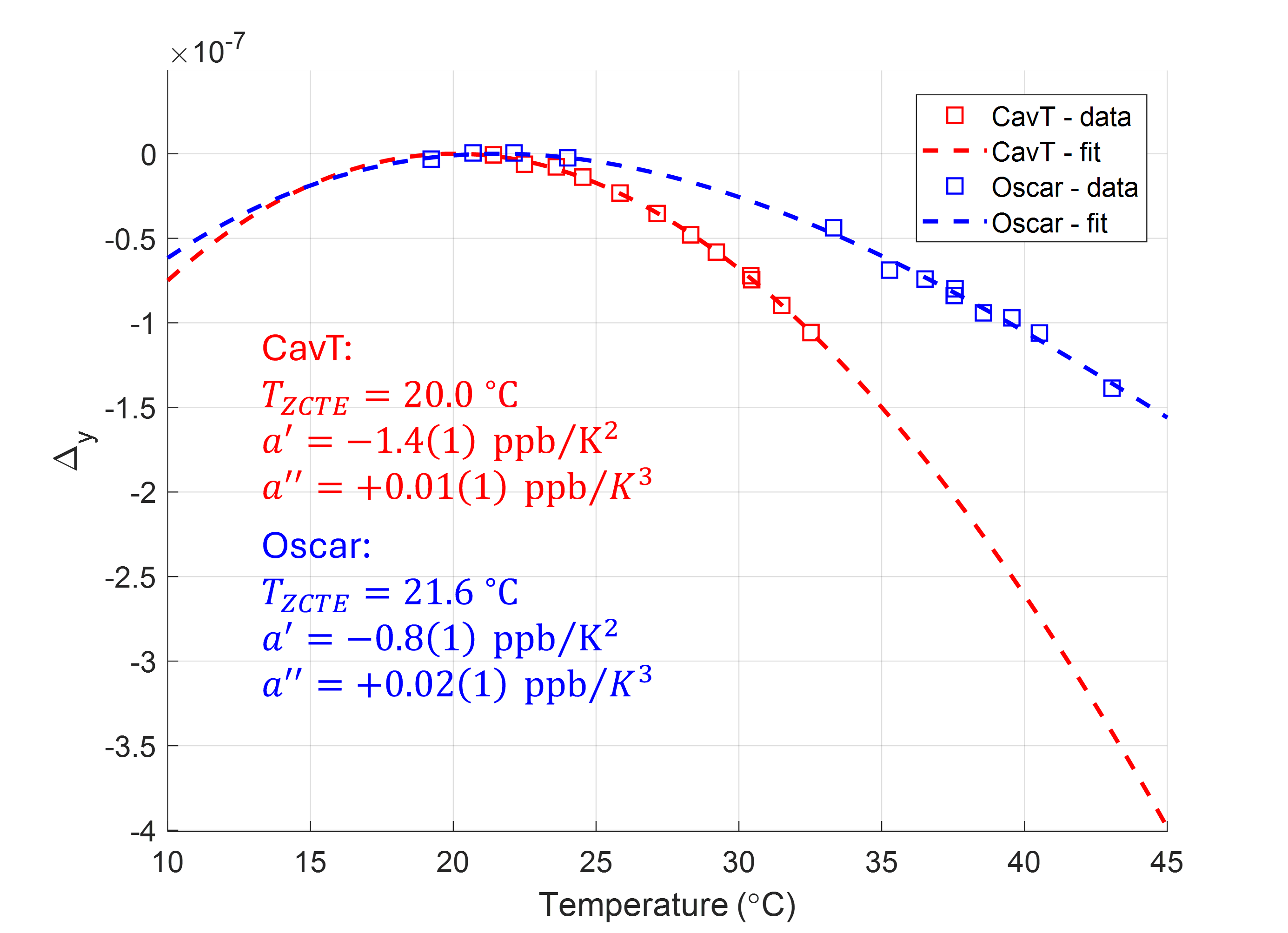}}
\caption{Thermal expansion of CavT and its precursor Oscar \cite{Both25}.}
\label{fig:ThermalExpansion}
\end{figure}

As discussed in the main text, we designed CavT to have a zero-crossing in the coefficient of thermal expansion (ZCTE) at 30 $^\circ$C, but we measured the ZCTE at 20.0 $^\circ$C. We present the thermal expansion measurement of both CavT and its precursor \cite{Both25}, nicknamed Oscar, in Figure \ref{fig:ThermalExpansion}. Both cavities' spacers and backing rings were cut from the same boule. We requested 7973 ULE with a native ZCTE of $40$ $^\circ$C with a tolerance of $\pm5 ^\circ$C. 

The design of these cavities was informed by simulating thermal expansion of the cavity assembly with a finite element model. In this model, the thermal expansion of the ULE elements is given by 
\begin{equation}
\alpha_{ULE}(T) = a'_{ULE} (T-T_{ZCTE})
\label{eq:ULECTE}
\end{equation}
while the fused silica substrates are modeled by a constant CTE $\alpha_{FS} = 500$ $\mathrm{ppb/K}$. The simulation returns the fractional length change between cavity mirrors $\Delta y$, and we fit $\Delta y$ to the form
\begin{equation}
\Delta y = \frac{1}{2} a' (T-T_{ZCTE})^2 + \frac{1}{3}a''(T - T_{ZCTE})^3
\label{eq:cavTEfit}
\end{equation}

The finite element calculations assumed the CTE slope of ULE to be $a'_{ULE} = 2.4$ $\mathrm{ppb/K^{2}}$, taken from \cite{Lege10}, which used a titania-doped glass of unknown origin. This reference over-estimates the measured CTE slopes we measured by a factor of 2 for CavT and a factor of 3 for Oscar, as shown in Figure \ref{fig:ThermalExpansion}. Corning 7973 ULE datasheet curves show a CTE slope of $1.7$ $\mathrm{ppb/K^{2}}$. However, our own measurements combined with recent ULE cavity literature \cite{Herb22,Krau25}, which also use Corning 7973 ULE, suggest a smaller typical value for the CTE slope of approximately $1.2$ $\mathrm{ppb/K^{2}}$. This discrepancy in the CTE slope can explain 4 $^\circ$C (9 $^\circ$C) of ZCTE discrepancy for CavT (Oscar), out of 10 $^\circ$C (8 $^\circ$C) discrepancy in total. The remaining discrepancy can be attributed to the $\pm5$ $^\circ$C tolerance of the native ZCTE of the ULE.

\section{Optimizing Cavity Photoresponse}

We found that one of the leading sources of technical noise on CavT was the response of the system to changes in optical power, as shown in Figure 2. Kraus et al. and Ma et al. have demonstrated a novel technique for compensating the photo-response of a cavity with $\mathrm{Al_{0.92}Ga_{0.08}As}$:$\mathrm{GaAs}$ mirrors by optimal choice of polarization axis and intracavity intensity \cite{Krau25,Ma26}. Due to scheduling pressure, we deployed the system to Mt. Blue Sky in 2025 before we could implement this technique on CavT. In the main text, the technical instability budget in Figure 2 and the stability measurements in Figure 6 labeled "Mountain 2025", reflect CavT's performance before we optimized CavT's photo-response.

\begin{figure}[htbp]
\centering
\fbox{\includegraphics[width=0.99\linewidth]{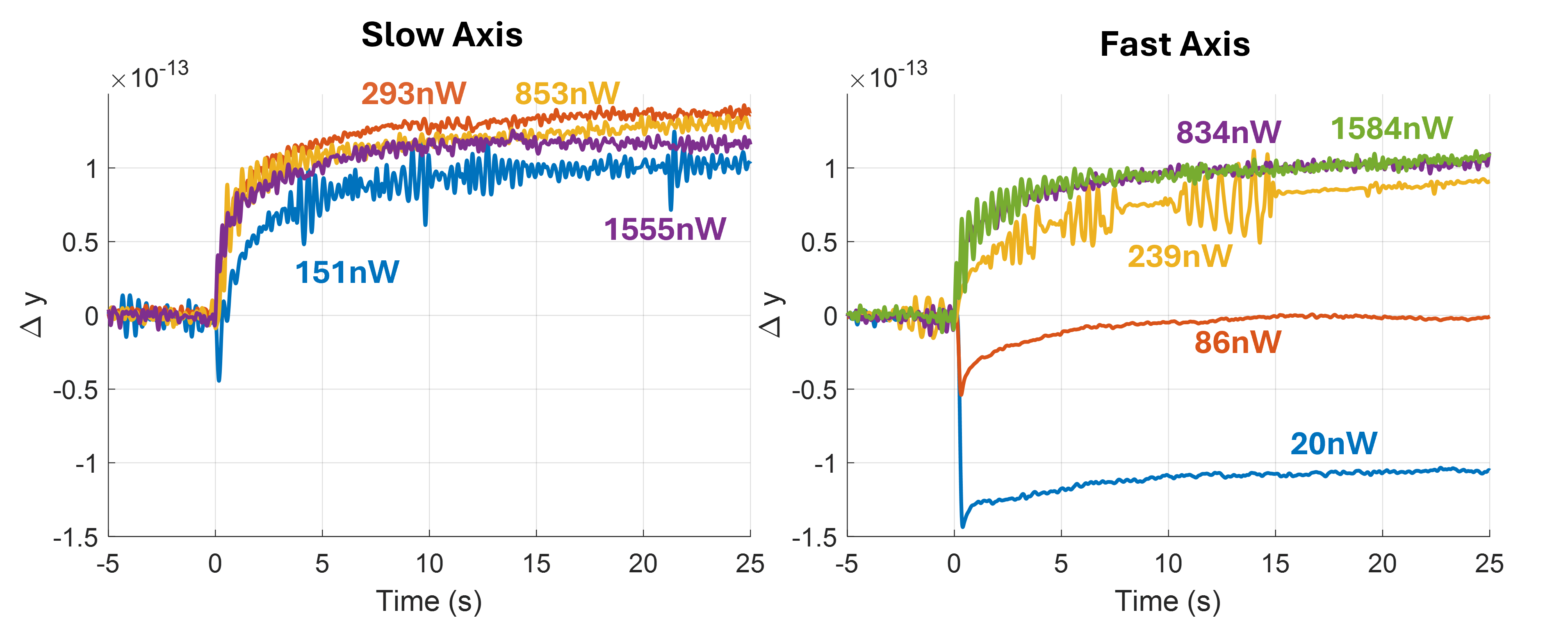}}
\caption{CavT Photoresponse versus polarization eigenmode and transmitted optical power. The transmitted optical power is written color-coded next to each corresponding measurement. Left: photo-response when the polarization is aligned to the slow axis of CavT. Right: photo-response when the polarization is aligned to the fast axis of CavT.}
\label{fig:photoresponse}
\end{figure}

Upon returning from Mt Blue Sky, we measured the response of CavT to a step change in optical power as a function of transmitted optical power and polarization axis (Figure \ref{fig:photoresponse}). We observed strong signatures of photo-birefringence, as shown in \cite{Krau25,Ma26}; however, we observed partial cancellation of photo-birefringent and photo-thermal effects on the fast axis. For comparison, Kraus et al. observed cancellation on the slow axis for the 20 cm transportable cavity \cite{Krau25}, but Ma et al. found cancellation on the fast axis for a 48 cm lab-based cavity \cite{Ma24}. While Equation 6 of \cite{Ma26} provide a detailed model of the time-dependence of the photo-birefringent effect, we find that a simpler two-exponential transfer function fits well to our measured photo-response and is adequate for evaluating the cavity's intensity budget. CavT deployed to Mt Blue-Sky with light coupled to the slow axis with a transmitted power of 690 nW, with a low frequency photo-sensitivity of 1.4 kHz/$\mu$W with time constants of 7.3 s and 0.42 s. We find the optimal configuration to be coupled to the fast axis with a transmitted power of 86 nW, with a low frequency photo-sensitivity of 0.3 Hz/$\mu$W with time constants of 3.5 s and 0.16 s.

\begin{figure}[h]
\centering
\fbox{\includegraphics[width=0.8\linewidth]{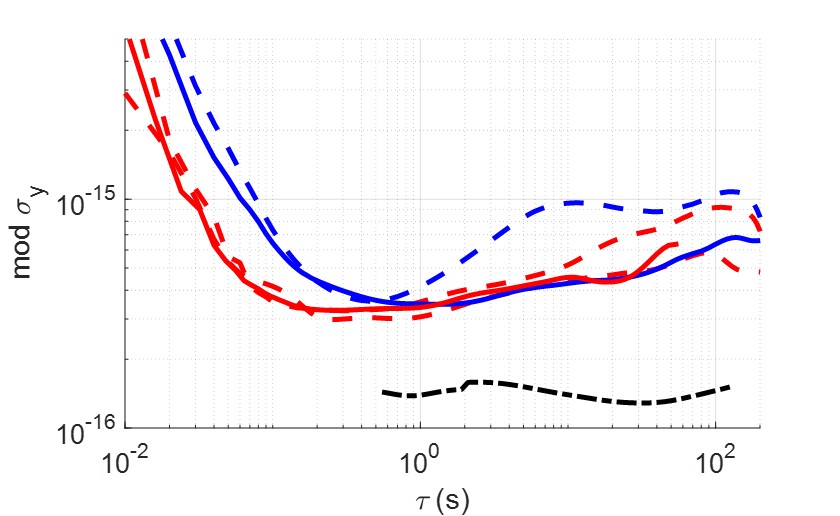}}
\caption{TSLS stability measurements over time against a 29 cm ULE reference cavity at NIST. Dashed lines indicate the cavity was coupled with slow axis light, while solid lines indicate the cavity was coupled with fast axis light with optimized photo-response. The red lines were stability measurements taken when the TSLS is in the lab at NIST, comparing with the reference cavity over ~10 m fiber. The blue lines are measured while the TSLS was deployed at the University of Colorado at Boulder, comparing to the same reference laser (located at NIST) over a 3.2 km fiber link as part of the Boulder Research and Administrative Network (BRAN). The stability of the reference laser is plotted in the dashed black line.}
\label{fig:stability}
\end{figure}

Figure \ref{fig:stability} summarizes the stability of CavT against a reference laser at NIST before and after we implemented partial photo-response cancellation and through two deployments. We implemented the partial photo-response cancellation during a deployment at the University of Colorado at Boulder. While there is some variation in the stability of CavT coupled to the slow axis for $\tau > 1$ s, we have yet to see significant variation in CavT stability when coupled to the fast axis with partial photo-response cancellation. The deterioration of stability for $\tau < 1$ s while the system was at the University of Colorado at Boulder is likely due to excess noise added by frequency transfer over 3.2 km of fiber between NIST and the university.

\section{Other Changes for the 2026 Deployment to Mount Blue Sky}
In addition to optimizing CavT's photoresponse, several other changes were made to the TSLS before or during we deployed to Mount Blue Sky in 2026. 

\subsection{Mitigating Thermal Gradient Fluctuations in CLaRC}
We observed that, once every 22.5 hours, the frequency of the 578 nm "Clock" FNC link would deviate to a peak value of $1\times10^{-17}$ then return to its original value within roughly 10 minutes, as observed in the heterodyne beat between the CLaRC 578 nm link and the lab-based laser 578 nm link, as in Figure 4 of the main text. This behavior is the origin of the bump near $\tau$ = 1000 s in the out-of-loop stability reported for the "Clock" link in Figure 4 in the main text. The origin of this highly repeatable phenomenon was discovered to be a daily thermal event originating from an uninterruptible power supply (UPS) located just below the CLaRC enclosure on the TSLS server rack (visible in Figure 1 in the main text). During this event, we observed a decrease in the heating power required to stabilize the temperature of the CLaRC baseplate, indicating a transient external heat load with a peak value of 20 mW. This transient heat load would cause the front of the CLaRC baseplate to rise by roughly 30mK and the back of the baseplate to fall by roughly 30mK. The largest lengths of uncompensated optical fiber between CLaRC's FNC links are the input and output fibers of the PPLN doubler. The PPLN is located in the front of CLaRC and thus experiences a local temperature change with the transient gradient. We eliminated this specific phenomenon by relocating the UPS away from CLaRC. To prevent any other source of transient thermal gradients from coupling into the "Clock" FNC link, we elevated the PPLN input and output fibers onto a small platform with its own Peltier elements for temperature control independent from the baseplate.

\subsection{FPGA-Based PDH Servo}
During the 2025 Mount Blue-Sky deployment, the fiber laser was locked to CavT using a two-part analog controller with fast and slow outputs, which control frequency modulation (FM) on a feedback fiber AOM at the output of the fiber laser (not shown in Figure 1 of the main text) and a piezotransducer on the fiber laser's resonator. This implementation was hindered by relatively frequent PDH unlocks that could only be resolved by adjusting the input offset of the fast controller. These issues were resolved after we switched the analog controllers for an digital controller built around a commercial FPGA platform at its core, similar to \cite{Cole24}. The PDH error signal is read at a 14 bit analog-digital-converter (ADC) at 250 megasamples per second. The ADC input is averaged over many samples to increase the resolution to enable splitting the cavity line at the part-per-million level. The digitized error signal is then fed into a PI$^2$D controller acting as the fast feedback stage. This fast controller modulates the frequency of on an on-chip direct digital synthesizer (DDS) driving the feedback AOM. The output of the fast PI$^2$D stage is also fed to a stage of PI$^2$ control (acting as the slow feedback stage), which is then fed to a 16 bit digital-to-analog converter (DAC) at 250 megasamples per second to generate the control voltage for the piezotransducer. We observed no statistically significant degradation in the stability of the TSLS after making this substitution, and the time between PDH unlocks improved from a few days to over a week.

\subsection{CavT Active RAM Stabilization}
The RAM instability shown in the right panel of Figure 2 represents the RAM measured out-of-loop on the PDH detector with the light was detuned off cavity resonance while there was no active RAM control and while the TSLS was still located in the temperature-controlled lab. During the 2026 Mount Blue Sky deployment, the TSLS instability was observed to be worse than during the 2025 deployment. An active RAM servo was implemented as an ad hoc solution, and anecdotally it seemed to improve the TSLS stability at short times. The stability data in the right panel of Figure 6 for the 2026 deployment was taken with the active RAM control engaged. The architecture of the active RAM control is based on \cite{Wong85}.

\bibliographystyle{plain}
\bibliography{ULETPaperBib,ULETPaperBibSI}

\end{document}